# Revealing the Predictive Power of Neural Operators for Strain Evolution in Digital Composites

*Meer Mehran Rashid[1], Souvik Chakraborty[2,3,*], N. M. Anoop Krishnan[1,3,4*]*

[1]Department of Civil Engineering, Indian Institute of Technology Delhi, Hauz Khas, New Delhi 110016, India.

[2]Department of Applied Mechanics, Indian Institute of Technology Delhi, Hauz Khas, New Delhi 110016, India.

[3]Yardi School of Artificial Intelligence, Indian Institute of Technology Delhi, Hauz Khas, New Delhi 110016, India.

[*]Corresponding authors: S. Chakraborty (souvik@iitd.ac.in), N. M. A. Krishnan (krishnan@iitd.ac.in)

**Abstract**

The demand for high-performance materials, along with advanced synthesis technologies such as additive manufacturing and 3D printing, has spurred the development of hierarchical composites with superior properties. However, computational modelling of such composites using physics-based solvers, while enabling the discovery of optimal microstructures, have prohibitively high computational cost hindering their practical application. To this extent, we show that Neural Operators (NOs) can be used to learn and predict the strain evolution in 2D digital composites. Specifically, we consider three architectures, namely, Fourier NO (FNO), Wavelet NO (WNO), and Multi-wavelet NO (MWT). We demonstrate that by providing a few initial strain frames as input, NOs can accurately predict multiple future time steps in an extremely data-efficient fashion, especially WNO. Further, once trained, NOs forecast the strain trajectories for completely unseen boundary conditions. Among NOs, only FNO offers super-resolution capabilities for estimating strains at multiple length scales, which can provide higher material and pixel-wise resolution. We also show that NOs can generalize to arbitrary geometries with finer domain resolution without the need for additional training. Based on all the results presented, we note that the FNO exhibits the best performance among the NOs, while also giving minimum inference time that is almost three orders magnitude lower than the conventional finite element solutions. Thus, FNOs can be used as a surrogate for accelerated simulation of the strain evolution in complex microstructures toward designing the next composite materials.

## 1. Introduction:

In response to the growing need for high-performance materials, it is imperative to investigate materials with a wide range of desired properties. In order to enhance the properties, we resort to composites that exhibit superior properties compared to their individual constituents (Barbero, 2017). Many such composites have already been practically realized with demonstrated superior performance (Chen et al., 2012; Kadic et al., 2019; Liu and Zhang, 2011, 2011; Meza et al., 2014; Pham et al., 2019; Su et al., 2020; Wegst et al., 2015; Zhang et al., 2020). However, conventional manufacturing methods have limited capabilities to tailor the material microstructure owing to challenges in combining the constituents in targeted fashion. The advent of additive manufacturing presents an opportunity for exploring countless design configurations that could lead to the development of advanced materials with unprecedented properties(Gu et al., 2016).

To comprehensively understand material behavior, conducting experiments for every possible configuration is impractical. Physics-based simulations such as finite element (FE) (Gu et al., 2016) (FE) and molecular dynamics (Chawla and Sharma, 2018; Kairn et al., 2005) (MD) have been widely adopted for modelling material behavior. However, these methods, aimed at full-field simulations, often turn out to be computationally expensive and time-consuming. Further, the design landscape in terms of the configurational space is huge, with limitless combinations, making it challenging to explore such space using existing solvers. With the growing popularity of machine learning (ML), new methods have emerged that overcome the limitations of such physics-based solvers. These ML methods can accurately predict physical fields with minimal training data without knowledge of the underlying mechanics. Unlike traditional solvers, ML methods offer transferability, allowing for the transfer of information across solution predictions.

ML models have shown remarkable performance in modelling complex physical phenomena. Their successful applications have been observed in various fields, including robotics, materials, and mechanics, among others. By exploiting large datasets, these models capture the underlying functions or the operators mapping the input to the output. One of the prominent applications of ML models is in learning the underlying partial differential equations (PDEs) (G. Gupta et al., 2022; Li et al., 2021; Tripura and Chakraborty, 2023) that govern the phenomena using purely data-driven methods. These models offer tremendous computational efficiency compared to existing methods. Recently, physics-informed neural networks (Raissi et al., 2019) (PINNs) have been introduced, reducing data dependence and ensuring that the models adhere to the underlying governing physical equations. In practice, a hybrid approach is often employed, which combines both data-driven and physics-informed methods to develop robust and accurate models. These models can be successfully used to solve continuum mechanics problems that are governed by PDEs and ODEs, which are otherwise resource intensive.

Machine learning (ML) methods have revolutionized the field of materials, enabling faster discovery of new materials, advanced material design, modelling, and optimization (Butler et al., 2018; Hughes et al., 2019; Jensen et al., 2019; Qin et al., 2020). A wide range of ML frameworks such as multi-layer perceptron, graph neural networks, convolutional neural networks, generative models, physics-informed neural networks, and neural operators have been utilized for characterizing materials and estimating their response and properties. (Cao et al., 2023) introduced a masked-fusion ANNs to model the j-shaped stress-strain curve in soft network materials. (Hasan et al., 2023) employed a microstructural fingerprint to predict the critical fracture stress levels. In a reduced model approach, authors used Gaussian Process Regression, random forests, and multi-layer perceptron to predict the critical fracture state. While analyzing the deformations in short fiber composites, (Friemann et al., 2023) employed a recurrent neural network (RNN) that predicted the path-depend elastoplastic stress response of short fibre reinforced composites. (Mozaffar et al., 2019) used RNNs to predict the plastic behavior of the composite representative volume element.

The use of convolutional neural networks (CNNs) has been widespread in predicting various material properties based on image data. CNNs have been a natural choice for such tasks due to their ability to detect local and global features from the image data (LeCun et al., 2015). (Nie et al., 2019) combined two CNNs to encode the structure, boundary conditions (BCs), and external forces of a cantilever structure to predict von Mises stress fields. (Yang et al., 2020) integrated Principal Component Analysis with CNNs to measure the stress-strain response over the entire failure path. (Bhaduri et al., 2022)utilized the UNet architecture to predict von Mises stress fields by inputting fiber configurations. In another study (A. Gupta et al., 2022), they proposed an accelerated multi-scale modelling approach using a similar architecture. Further, PINNs, that enforce the underlying governing laws by minimizing the residues of the equations, has also been utilized to solve the forward and inverse problems involving material mechanics.(Henkes et al., 2022) employed PINNs to model micromechanics for linear elastic materials. (Haghighat et al., 2021)applied PINNs to explore linear elasticity and elastoplasticity by incorporating momentum balance and constitutive relations into the PINNS. Similarly, (Zhang et al., 2022) formulated a PINN-based framework for identifying unknown geometric and material parameters. (Goswami et al., 2022) used a physics-informed variational formulation of DeepONet to predict the fracture path in quasi-brittle materials.

Several investigations have employed graph neural networks (GNNs) to describe and design intricate materials using a graph representation of their microstructural data. To model the dynamics of metamaterials, (Xue et al., 2023) used a metamaterial GNN based on a lattice representation of their structure. (Guo and Buehler, 2020) utilized a GNN based on a semi-supervised approach to design architected materials. More recently, (Maurizi et al., 2022) utilized GNN frameworks to estimate stress, strain, and deformation maps for various material systems. Numerous research works have effectively utilized generative models to solve an inverse problem to attain a specific material possessing a target property. (Yang et al., 2021a)utilized conditional Generative Adversarial Networks (cGANs) to model the constitutive relationship by estimating the stress-strain fields, obtained by training the model on microstructure images. Another study conducted by (Yang et al., 2021b) trained multiple models to predict all components of the stress-strain tensor. (Rashid et al., 2022) used a single trained FNO model to obtain full stress and strain tensor besides demonstrating zero shot super-resolution capabilities. (Oommen et al., 2022)employed a blend of a convolutional autoencoder and DeepONet to predict microstructure evolution.

Most of the studies discussed above have primarily focused on predicting strain or strain components at a single instance of time. To completely mimic the FE solvers, it is desired to predict the deformation trajectory of the material and not just some individual time step. Besides, having information about every deformation time setup allows a better understanding of the material behavior. To address this, we employ neural operators (NOs) *viz* Fourier neural operator (FNO) (Li et al., 2021), wavelet neural operator (WNO) (Tripura and Chakraborty, 2023) and multi-wavelet neural operator (MWT) (G. Gupta et al., 2022) to predict the strain evolution of 2D digital composites. Our research showcases the NOs' potential to capture the strain fields for multiple future time steps with high accuracy based on a few initial strain field snapshots. Furthermore, we exploit the material and pixel-wise super-resolution features of NOs to obtain high-fidelity and high-resolution outcomes using a cost-effective trained model. Additionally, we validate the ability of NOs to generalize to unseen boundary conditions and predict the strain trajectory simulated for arbitrary microstructure geometry. Our research findings suggest that NOs can offer a robust framework as an alternative to high-fidelity solvers for the accurate prediction of strain field trajectory.

## 2. Methodology:

**Neural Operators:** Neural networks have been used to learn mappings between finite-dimensional Euclidean spaces. In such network architectures, discrete inputs are used to establish the underlying relationships via a standard supervised learning framework. In recent years, a novel paradigm called the neural operator (NO) has emerged to learn mappings between infinite-dimensional Euclidean spaces, which is a generalization of traditional neural networks (Li et al., 2020; Lu et al., 2021; Patel et al., 2021). This approach is helpful in learning the operator that maps the input function space to the solution space, allowing for more complex and flexible modelling of systems.

We use these neural operators to solve the PDEs by specifically learning the operator that maps the input parameters $a \in \mathcal{A}$ to the solution space $u \in \mathcal{U}$. Let $D \subset \mathbb{R}^d$ be bounded and open set and $\mathcal{A} = (D; \mathbb{R}^{d_a})$ is the input function space, $\mathcal{U} = (D; \mathbb{R}^{d_u})$ is the output function space. These spaces contain functions that map the domain D to $\mathbb{R}^{d_a}$ and $\mathbb{R}^{d_u}$ respectively. $\mathcal{G}: \mathcal{A} * \theta \mapsto \mathcal{U}$ is the mapping that satisfies the PDE. Considering samples $\{a_j, u_j\}$ where $a_j$ is an independent and identically distributed (i.i.d) sequence sampled from the probability measure $\mu$ in $\mathcal{A}$ and $u_j = \mathcal{G}(a_j)$, the neural operator approximates the mapping $\mathcal{G}_\theta$ by minimising the following stated problem using the cost function $C: \mathcal{U} \times \mathcal{U} \mapsto \mathbb{R}$

$$\min_\theta E_{a \sim \mu}[C(\mathcal{G}_\theta(a), \mathcal{G}(a))] \qquad \text{(Equation 1)}$$

Within the framework of the problem, we assume input function $a_j$ and solution function $u_j$ are evaluated point-wise. Let $D_j = \{x_1, x_2, \ldots, x_m\}$ be the $m$ point discretization and $a_j$ and $u_j$ be the finite samples of input-output pairs accessible. In this computational setup, we work with these finite $m$ pair data $\{a_j, u_j\}_{j=1}^{m}$ to learn the non-linear differential operator $\mathcal{G}_\theta$ which approximates the $\mathcal{G}: \mathcal{A} \mapsto \mathcal{U}$ satisfying the governing PDE. Here, we present a brief overview of the model architecture of the FNO, WNO, and MWT used for this study.

**FNO:** was introduced to solve the parametric partial differential equations. As shown in Figure 1, FNO consists of lifting layer P, Fourier Layers, and a projection layer. The input image is lifted to a higher dimension using P, which is a linear layer. The higher dimensional output goes iteratively through Fourier layers. Within each Fourier layer, the physical representation is supplied concurrently to the convolution and Fourier decomposition layers. The Fourier decomposition is performed using the FFT algorithm. The output is filtered by truncating the higher modes associated with spatial details while as the lower modes responsible for global features are retained. These filtered modes are transformed back to the spatial domain using inverse FFT. This output is summed with the convolution layer output before being fed to another Fourier layer. Finally, FNO uses a projection layer Q to transform the output to the target dimensions.

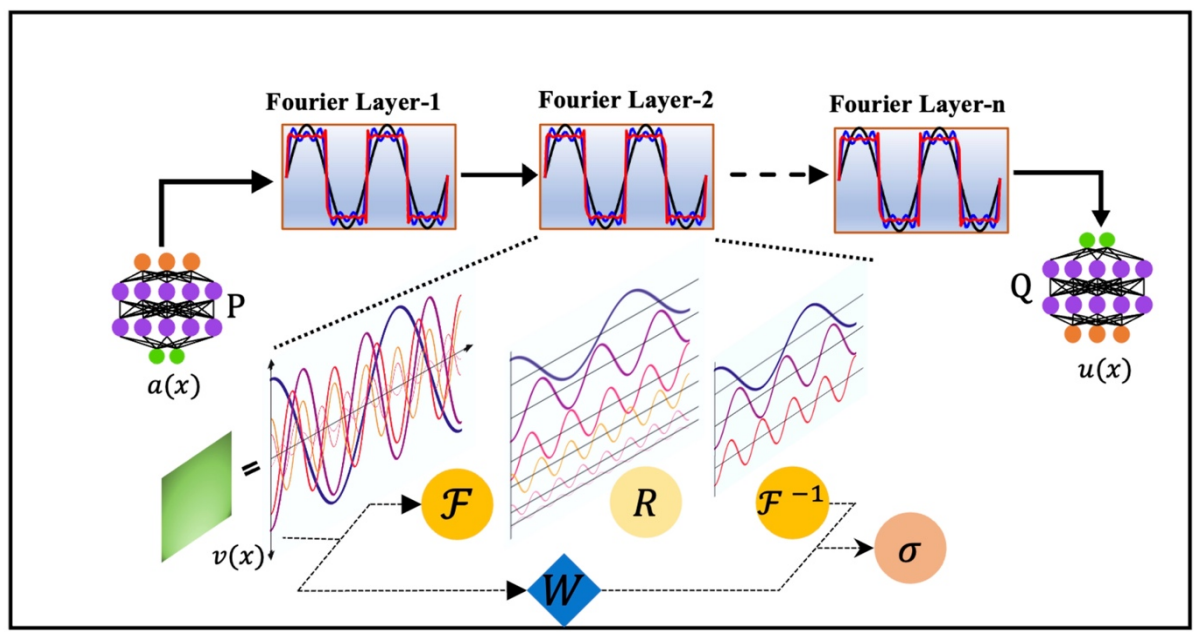

**Figure 1**. FNO architecture. The input image is lifted to a higher dimension using P, which is a linear layer. The higher dimensional output goes iteratively through Fourier layers. Within each Fourier layer, the physical representation is supplied concurrently to the convolution and Fourier decomposition layers. The output is filtered by truncating the higher modes associated with spatial details while the lower modes responsible for global features are retained. These filtered modes are transformed back to the spatial domain using inverse FFT. This output is summed with the convolution layer output before being fed to another Fourier layer. Finally, FNO uses a projection layer Q to transform the output to the target dimensions

**WNO:** Similar to FNO, WNO uses wavelets to decompose the input functions. The model performs wavelet transform on the images to get the sub-band coefficients. The images undergo recursive decompositions until the number of decomposition levels is exhausted. Finally, the sub-band coefficients of the last level are used for model parameterization. The coefficients of initial levels of decomposition contain coefficients associated with higher frequencies of the input image, while the last levels correspond to the low-frequency features or the global features. The level of decomposition is a hyperparameter. As shown in Figure 2, the input is lifted to a higher dimension using the local transformation P(.). The input passes through a series of wavelet kernel integral layers. Finally, the output is transformed to the target dimension using Q(.). Within each wavelet kernel layer, the input consists of spatial coordinates and input function. The output of P is fed concurrently to the wavelet decomposition layer and a 1D convolution layer. The output of both layers is summed and activated before feeding it to another wavelet kernel integral layer. After going through a series of such layers, the output is transformed to the target dimension using the projection layer Q. Both P and Q are essentially shallow neural Networks. The readers are advised to refer to the original work(Tripura and Chakraborty, 2023) for a detailed explanation.

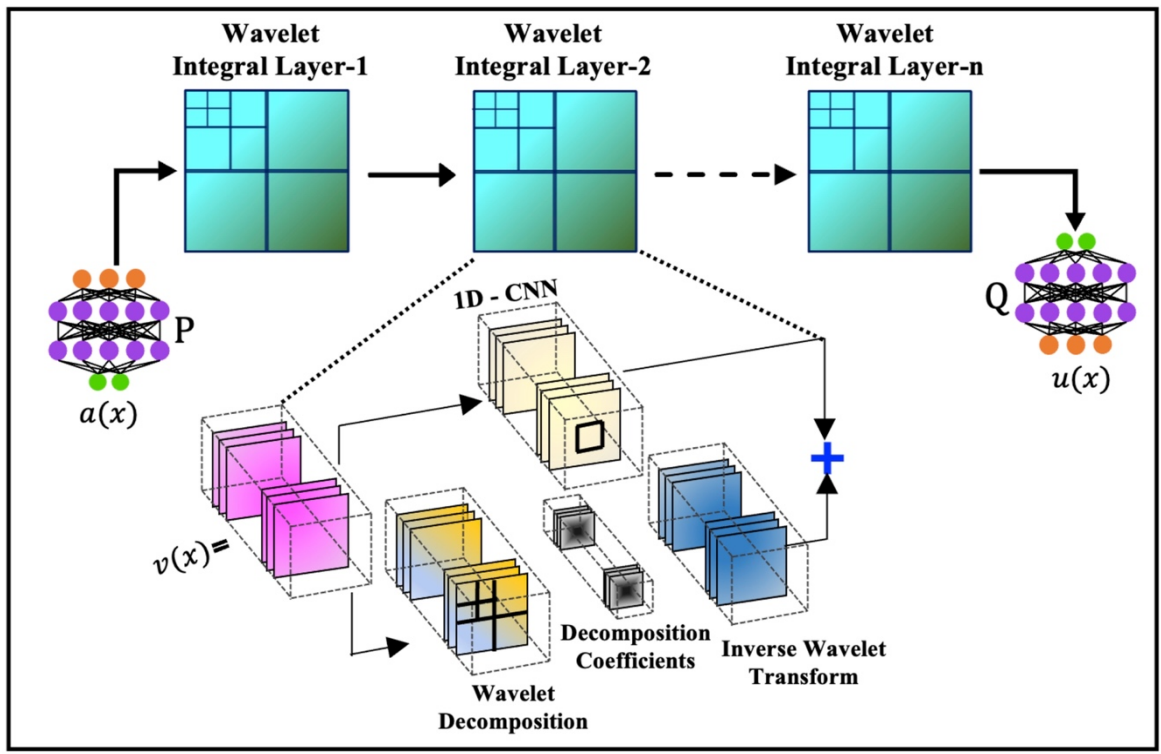

**Figure 2**. WNO Model Architecture. The input is lifted to a higher dimension using the local transformation P(.). The input passes through a series of wavelet kernel integral layers. Finally, the output is transformed to the target dimension using Q(.). Within each wavelet kernel layer, the input consists of spatial coordinates and input function. The output of P is fed concurrently to the wavelet decomposition layer and a 1D convolution layer. The output of both layers is summed and activated before feeding it to another wavelet kernel integral layer. After going through a series of such layers, the output is transformed to the target dimension using the projection layer Q.

**MWT:** composed of a recurrent unit that includes two distinct parts, namely the decomposition and reconstruction part as shown in Figure 3. The decomposition part uses the input data to calculate the multi-scale multiwavelet coefficients at a coarser level based on pre-computed transformations utilizing special orthogonal polynomials. These coefficients are then fed into four neural networks representing the kernel approximations. In the case of 2D data, the first three neural networks are composed of single-layer convolutional neural networks (CNNs) followed by a linear layer, while the fourth network is a single linear layer. The reconstruction step then utilizes the output of these NNs to calculate the coefficients for the finer scale. This process is repeated iteratively until the coefficients for the finest scale are achieved. For a detailed explanation of the method, the interested reader is referred to the original work(G. Gupta et al., 2022).

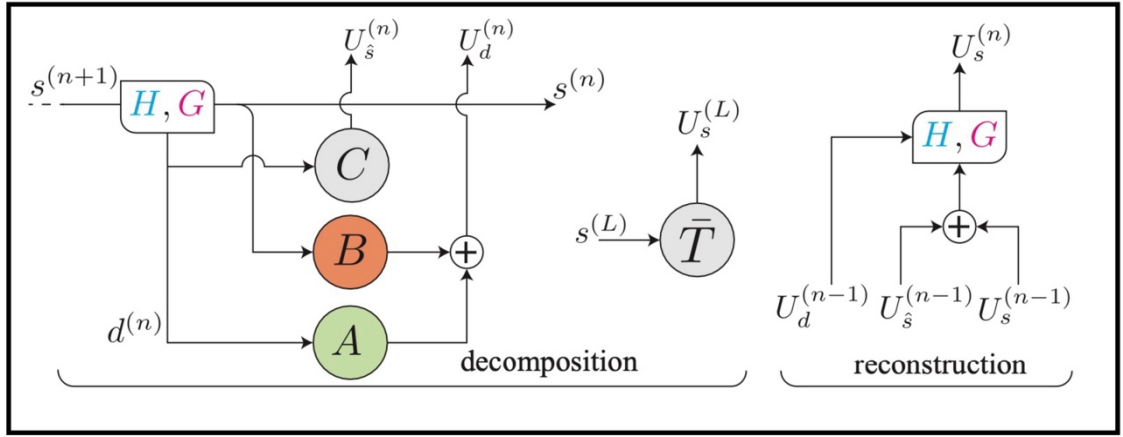

**Figure 3.** MWT architecture (G. Gupta et al., 2022). It is composed of a recurrent unit that includes two parts; decomposition and reconstruction part. The decomposition part uses the input data to calculate the multi-scale multiwavelet coefficients at a coarser level based on pre-computed transformations utilizing special orthogonal polynomials (H,G). These coefficients are then fed into four neural networks, A,B,C and T, representing kernel approximations. A,B and C are single-layer convolutional neural networks (CNNs) followed by a linear layer, while T is a single linear layer. The reconstruction step then utilizes the output of these NNs to calculate the coefficients for the finer scale. This process is repeated iteratively until the coefficients for the finest scale are achieved.

In order to train the neural operators (NOs), we adopt the teacher-forcing(Lamb et al., 2016) approach commonly employed in training recurrent neural networks (RNNs). This approach involves feeding the ground truth from a previous time step as input for the subsequent time step instead of using the model prediction as input. The use of teacher forcing is known to alleviate convergence issues and instability in the model's training. For all the NOs, we adopted a common approach of inputting spatial coordinates and strain snapshots at time steps t={$t_1,t_2,t_3$}. These models predict the strain at next time step, i.e, $t_4$. During testing, the output at time step $t_4$ is concatenated with the input while dropping t1. Thus we have [x,y,{ $t_1,t_2,t_3$}] as input for next iteration. This ensures that a fixed window of time steps (a window of 3 steps) is fed to predict the next time step until we obtain the predictions for all 10 time steps, up to

$t_{13}$. It is important to mention that NOs have no recurrent connections and, therefore, independent of temporal dependence both in the training as well as testing stage.

## 2.1. Dataset Preparation using Finite Element Modelling

To generate the initial dataset, we performed Mode-I tensile test FE simulations on an 8 mm × 8 mm 2D plate using ABAQUS (Manual, 2014) software. The thickness of the plate was considered negligible (0.001 mm) compared to the other dimensions, and was ignored for numerical computation purposes. The composite material used in the simulations consisted of two components: a soft and a stiff material. The modulus ratio ($E_{stiff}/E_{soft}$) for the two materials was 10, and the failure stress for both materials was arbitrarily set at 40 MPa. As a result, the toughness ratio ($G_{stiff}/G_{soft}$) of the two materials was 0.1. Both materials were assumed to be perfectly elastic, and the point of damage initiation was defined using a maximum principal stress criterion. Herein, we used a fracture energy-based damage evolution criterion to simulate the softening behavior of an arbitrary composite material under mode-I tensile testing. An 8 mm × 8 mm 2D square plate was divided into equal cells, and each cell was assigned a random material property (soft or stiff) using an in-house python script. However, the fraction of soft and stiff units remained the same for all FE samples, resulting in a material resolution of 8×8 and an overall image resolution of 48×48. Each pixel was represented as a FE in the configuration, and the loading was applied in the horizontal (global *y*) direction (see Figure 4), while the pre-crack was along the x-direction. We generated approximately 1000 distinct configurations with varied design arrangements and used them to simulate mode-I tensile failure of the composite plate. We also generated multiple test sets with different material and image resolutions, as discussed in the results section. Finally, we post-processed the results using ABAQUS's Python interface to extract strain trajectories for all the simulated samples.

## 2.2 Machine Learning Model Development

In training the NOs to solve the strain trajectory, we adopt the teacher-forcing approach that is commonly used for training RNNs, which leads to better convergence and training stability. The choice of the appropriate loss function is crucial in the training of NOs. High-gradient regions in the strain maps are usually difficult to capture for any deep-learning model. If a regular $L_2$-based loss function is used, the models prioritize minimizing the high error near the interface and crack tip regions. However, we aim to learn the entire map with local and global features. Therefore, adopt an $L_2$-based loss function described in Equation (3), which ensures low error near the high-gradient regions and higher accuracy for overall smoother features.

**2.3 Model Hyperparameters**

Although each NO has a distinct set of hyperparameters, we will discuss the shared ones here, and the others will be addressed later. All models were trained using the PyTorch (Paszke et al., 2019) framework on an NVIDIA A5000 GPU. We used ADAM optimizer, a first-order gradient-based method for training, with learning rate decay controlled by a *steplr* scheduler. The dataset was divided into 1000 training and 200 testing trajectories. We use a smoother version of the *ReLU*; The GELU (Gaussian cumulative distribution function) as the activation function for FNO and WNO.

**FNO** model architecture, illustrated in Figure 1, utilizes four Fourier layers, where each layer retains the first 12 Fourier modes and is then summed with the output of a 1D convolution layer. The lifting layer P and projection layer Q each have 32 nodes. The FNO model is trained for 500 epochs with a learning rate of 0.001, batch size of 20 and weight decay of 1e-3.

**WNO** architectural details are shown in Figure 2. The lifting layer P and Projection layer Q have 64 nodes each. There are a total of 4 kernel integral wavelet layers and within each such layer, we perform a level 1 decomposition and the convolution layer is one-dimensional. The model is trained for 1000 epochs with a learning rate of 0.0001, the batch size is 20 and weight decay of $1e^{-3}$.

**MWT**: the architecture is shown in Figure 3. The dec uses 4 neural networks A B C and T where in A, B and C represent a single CNN followed by Linear layer and T is a single linear layer. We use Legendre polynomials to calculate the multi-scale multiwavelet coefficients. The rec part uses the output of dec to get the finer multi-scale multiwavelet coefficients of the solution. Four such units of rec and dec are used in this study. The model uses ReLU as an activation function. MWT is trained for 500 epochs with a learning rate of 0.001 using a batch size 40 and weight decay of 1e-5.

**3.4 Evaluation Metrics**

In our study, we use Neural Operators to predict the strain snapshots at discrete time intervals, forming the strain trajectory together. Our goal is to achieve prediction results that are as accurate as those obtained from high-fidelity solvers. To evaluate the accuracy of our predictions, we adopt the AE metric, which measures the pixel-wise error for each timestep and is defined by equation 2. A lower AE value indicates higher accuracy of the model predictions.

$$AE := \delta_{AE} = |\hat{u}(x_i) - u(x_i)| \qquad \text{Equation 2}$$

## 3 Results and Discussion

### 3.1 ML model development

NOs have been used to solve many problems. Just like conventional neural nets, which map the finite input vector the finite output vector, NOs map input functions to the solution functions by approximating the underlying operator. Once effectively trained, these models exhibit discretization invariance, i.e., they share common network parameters irrespective of the parameterization used for the underlying input function. The NOs have been shown superior capabilities in learning the underlying partial differential equations to provide accurate inference without the use an explicit solver. The high-level workflow employed in this study is shown in Figure 4. First, we generate many random microstructures with a chequerboard type geometry of two-phase composites subjected to tensile loading using classical FE modelling in a quasi-static fashion. The strain evolution of the composite for several steps is captured while ensuring that the composite does not initiate any cracking (in the extended FE framework). The dataset is then used to train the NOs, namely, FNO, WNO, and MWT to predict the strain evolution in these composites. The trained models can then be evaluated on unseen microstructures, new boundary conditions, or different resolutions to evaluate their capabilities to predict the strain evolution.

For a 2D composite featuring an 8 × 8 material chequerboard configuration, the number of potential discrete geometries is $2^{64}$. Applying conventional FE techniques to address such a vast array of geometries would be impractical and computationally prohibitive. Thus, we estimate the time series trajectory of the composites using FNO, WNO and MWT. Here, the input tensor typically takes the form of a tensor with dimensions $\{x, y, [T = 0:t]\}$, where, $x, y$ are the spatial coordinates, and $T$ denotes the number of time steps. Further, Nos are used to forecast the structures future behavior by predicting a sequence of subsequent time $t_n$, where $n$ represents the number of future time steps to be predicted. The ultimate objective is to obtain predictions of multiple future time steps within the acceptable error tolerance.

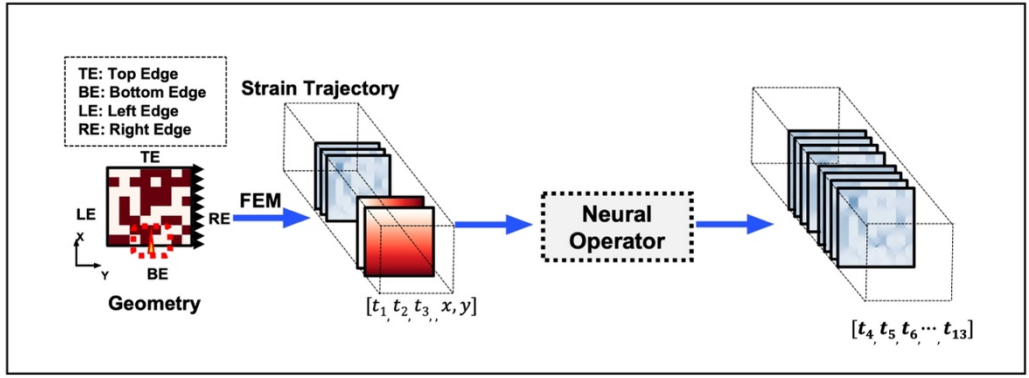

**Figure 4**. Chequerboard type geometry of two-phase composites is analysed under mode-1 tensile failure. The load is applied in the y-direction. The strain trajectory consisting of strain values in the y direction are fetched along with element coordinates. This data is supplied to the NOs, viz, FNO, WNO, and MWT to train the models so as to predict the future strain values of the composite.

## 3.2. Strain Evolution

Figure 5 presents a comparison of predictions obtained from NO framework and high-fidelity FE simulations for a representative case of strain trajectory evolution at three distinct time steps. The input to the model is a $48 \times 48 \times 5$ strain tensor, having the strain information for the first three time steps and grid information. The overall output from the model is a tensor of dimensions of $48 \times 48 \times 10$ where 10 denotes the time steps from $t_4$ to $t_{13}$. Here, for the sake of brevity, we report the results for t = 4, 9 and $13^{th}$ time step. It is expected that the data-driven approaches accumulate error when predicting the forward trajectory. Accordingly, the performance for the last timestep, that is, $13^{th}$ provides the upper limit of the error in the predicted trajectory. As evident from the Figure 5, prediction maps generated by NOs exhibit strong agreement with the ground truth results in both qualitative and quantitative terms. However, small discrepancies are observed in the interfaces between the soft and stiff regions, particularly for MWT. During the later time steps, the strains develop rapidly due to the imminent crack propagation, and hence the accuracy drops relatively for predicted strain towards the later steps. For a better understanding of the accuracy of the output, we calculate the pixel-wise absolute errors (AE) calculated as per Eq. (2), which are plotted for every time step. Pixels around the crack tip and interphase regions report relatively higher errors due to the development of localized stress concentrations. These errors suggest that the models are unable to precisely capture the spatial gradient of the strain fields in the vicinity of the interface. For the rest of the material domain, the error plots show that Nos can exactly match the ground truth with acceptable errors. As expected, the magnitude of the error increases as the time step increases. This is normal for most of the time series ML models, wherein the solution diverges as the number of predicted time steps increase. The ability of the Nos to predict strain evolution is impressive, especially the intricate details near the phase interface and crack tip. These results demonstrate the feasibility of NOs in replacing the existing physics-based solvers saving time by orders of magnitude.

While training the models, we varied *n* (the number of time steps) predicted from 1 to 12. The results are shown in Figure 6(a). The solutions significantly diverge after ~$14^{th}$ timestep, as the crack propagation begins. The figure shows the error for every *n* strain steps predicted up to 14 for the corresponding NOs. Previous studies have primarily focused on mapping input geometries to specific time steps in stress or strain trajectories. In contrast, this study prioritizes predicting multiple time steps through a single forward pass using a pre-trained model. However, it should be noted that predicting more time steps into the future leads to decreased accuracy. As seen from the Figure 6(a), with increasing timesteps in the output, the models' performance decreases. It is noteworthy that when predicting for t > 14 steps, the predictions deviate significantly for almost all the time steps. During training, the loss propagated is calculated as the total loss for all the time steps. This becomes problematic for crack/fracture propagation as these models struggle to capture the discontinues in the field quantities. As a result, for time steps that characterize crack propagation, the losses are high, which in turn increases the total loss. This results in the model prioritizing the minimization of the error for the last few time steps to reduce the overall error, resulting in inaccurate predictions for a longer time horizon. Therefore, for this study, we have predicted the strains from t = 3 to 14, which where the crack propagation has not yet started for any of the composite geometries considered.

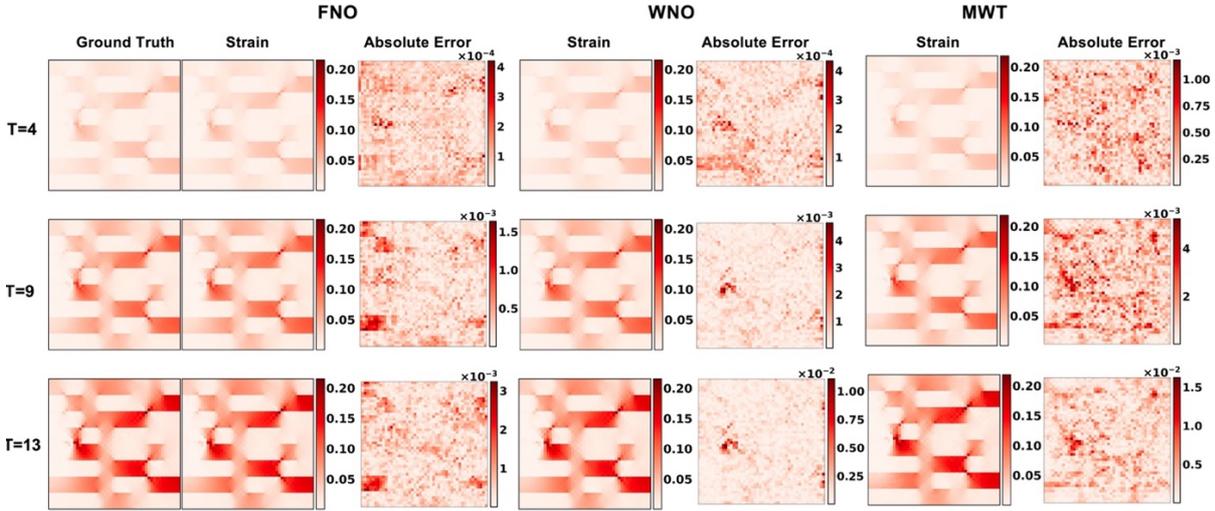

**Figure 5.** Strain Evolution for FNO, WNO and MWT for time, T = 4, 9, and 13 steps, compared to ground truth. The ground truth given by FEM, predicted strain by the NO architectures and the respective absolute errors are shown.

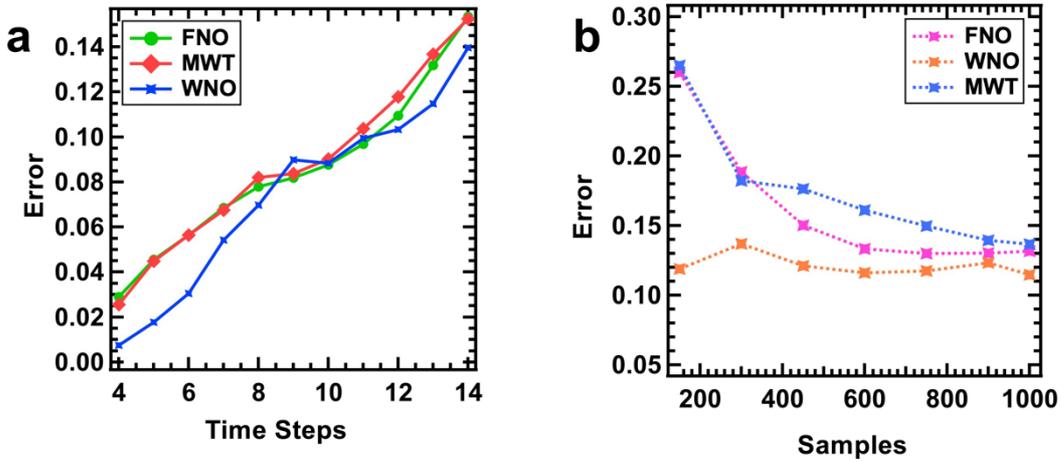

**Figure 6. (a)** Effect of number of output steps on the error in the predicted strain by FNO, MWT, and WNO. **(b)** Effect of total training samples on the performance of models. We note that WNO is highly data efficient.

To assess the precision of our outcomes, we opt to examine two slices XX and YY, across the horizontal and vertical dimensions of the 2D composite We plot the strain variation along these slices in Figure 7 for the NOs. The lines represent the strain evolution from t = 4 to t=13 predicted by the NOs. The strain variation plots indicate that the NOs can effectively estimate the nonlinear behavior of the strains, and the predicted results exhibit an excellent match with the ground truth, even at the phase interface. The MWT shows a weaker performance in predicting the last time step compared to FNO and WNO, which exactly match the FE results. The high accuracy achieved by NOs allows for capturing the complex behavior of strain fields, making them a useful tool for high-fidelity analysis, facilitating the exploration of the design landscape and enhancing our understanding of material behavior. In order to evaluate the comparative performance of the models, we present the overall test results for in Table 1. The errors are computed according to Equation 3, and the results are compared against a test set consisting of 200 strain trajectories. The results show that the performance of all three models is comparable, with WNO exhibiting the least error in relation to the other models.

$$\mathcal{L}_2 = \frac{\sqrt{\sum_{i=1}^{m}(u(x_i) - \hat{u}(x_i))^2}}{\sqrt{\sum_{i=1}^{m}(u(x_i))^2}} \qquad \text{Equation (3)}$$

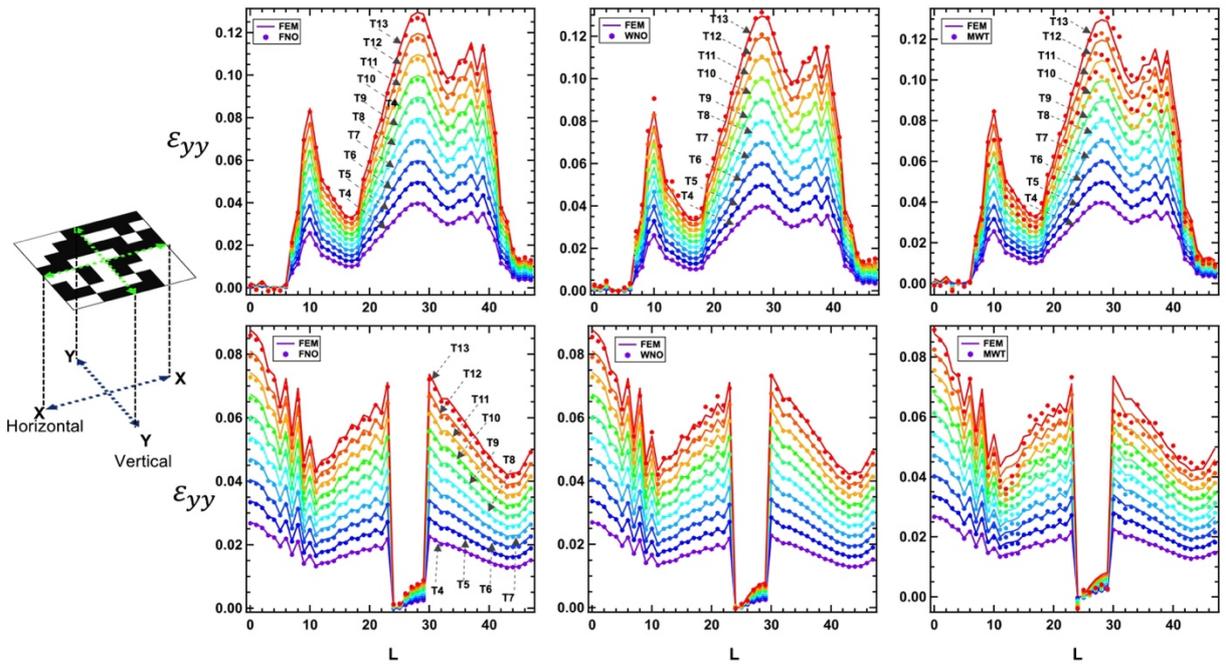

**Figure 7**. Strain values along horizontal and vertical directions. Quantitative comparison of strains for FNO, WNO and MWT along the **(a)** horizontal direction (XX), and **(b)** vertical direction (YY).

In the context of training deep learning models, a key consideration is the amount of training data required to achieve a reasonable level of accuracy. Prior studies demonstrating the prediction of physical field variables, such as stress or strain, often rely on training models using few thousand images. However, the cost and computational resources required to train models on large datasets can be prohibitive. In Figure 6(b), we present the performance of NOs as the size of the dataset is decreased. Notably, even with very small datasets, the models can achieve high accuracy.

Table 1 presents the total number of trainable parameters and the corresponding training time for each model. In many cases, training time is not given sufficient consideration when evaluating model performance. Even though WNO has the lowest test error as shown in Table 1, the training time for WNO is considerably longer than the other models. The superior performance of WNO can be attributed to its higher number of model parameters. However, this also leads to increased expenses during the training process.. The test set used for all datasets is the same, and the error reported is calculated as per Equation (3). These findings demonstrate that NOs are not data-hungry and can accurately predict highly nonlinear material responses using a small dataset.

| Model | Error | | Time | | Parameters |
| --- | --- | --- | --- | --- | --- |
| | Strains | BCs | Training (mins) | Inference (sec) | |
| FNO | 0.1318 | 0.1139 | 49 | 0.29 | ~$2.1 \times 10^6$ |
| WNO | 0.1147 | 0.1315 | 104 | 1.73 | ~$7.1 \times 10^7$ |
| MWT | 0.1368 | 0.0488 | 404 | 2.41 | ~$1.8 \times 10^6$ |

**Table 1. Performance Metrics**. The test set has total 200 sample trajectories and the results are shown in Strains column. The overall test results for all the boundary conditions is in BCs. We log the training and inference time as well as the model size in terms of total trainable parameters.

### 3.3 Multiple Boundary Conditions

Until now, the results presented in this study are based on various simulated strain trajectories obtained from the analysis of diverse 2D composite geometries with identical boundary conditions, as depicted in Figure 8. However, in practical applications, materials may experience varying boundary conditions. One approach to predict such conditions is to train the ML model using a diverse dataset that includes strain trajectories corresponding to various boundary conditions. Nonetheless, this leads to an increase in the size of the dataset, which in turn extends the training time. Here, we show the flexibility of NOs to predict strain evolution for different boundary conditions (BCs). Although the models are trained on a single BC, they are able to predict strain evolution for unseen BCs. To test the performance of the NOs, we created a test set consisting of 12 different BC cases, as shown in Table 2. Figure 8 presents the results for a typical BC case, thereby demonstrating the zero-shot generalization ability of the NOs. Moreover, the overall error on all cases for all NOs is reported in Table 1 in the BCs column. The findings suggest that NOs can predict strain evolution for different boundary conditions, potentially reducing the training time required for training the model on different BCs. Amongst the three NOs, MWT has the least error, whereas FNO and WNO show a similar performance. The ability of NOs to generalize to different boundary conditions offers significant advantages, allowing for precise results for any boundary case without retraining the model for each case.

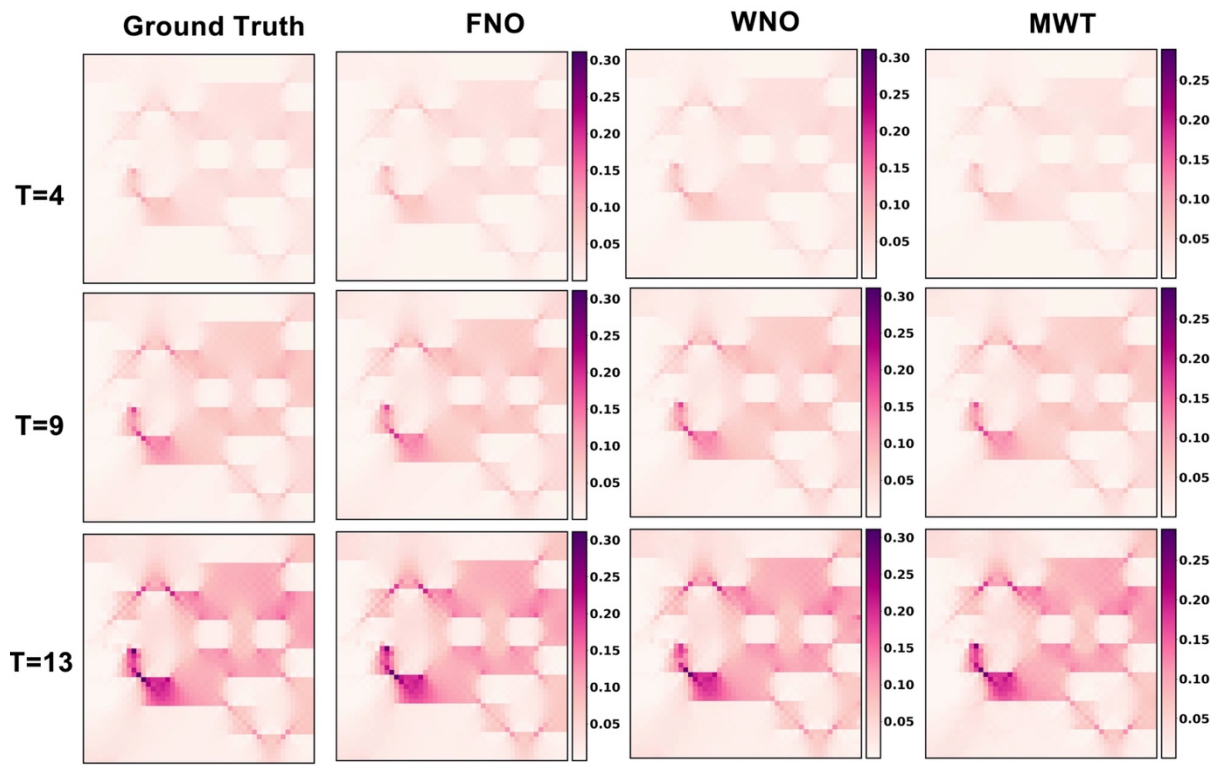

**Figure 8.** The model trained on a specific boundary condition is used to predict for unseen boundary conditions. Ground truth and predictions of FNO, WNO, and MWT at three different times, namely, T = 4, 9, and 13.

| Case | Left Edge (LE) | Right Edge (RE) | Bottom Edge (BE) |
|---|---|---|---|
| Training | U1 = 0, U2 = Free, UR3 = 0 | U1 = 0, U2 = Free, UR3 = 0 | U1 = 0, U2 = 0, UR3 = 0 |
| Case 1 | U1 = 0, U2 = Free, UR3 = 0 | U1 = 0, U2 = Free, UR3 = 0 | U1 = 0, U2 = 0, UR3 = Free |
| Case 2 | U1 = 0, U2 = Free, UR3 = Free | U1 = 0, U2 = Free, UR3 = 0 | U1 = 0, U2 = 0, UR3 = 0 |
| Case 3 | U1 = 0, U2 = Free, UR3 = 0 | U1 = 0, U2 = Free, UR3 = Free | U1 = 0, U2 = 0, UR3 = 0 |
| Case 4 | U1 = 0, U2 = Free, UR3 = 0 | U1 = 0, U2 = Free, UR3 = 0 | U1 = Free, U2 = 0, UR3 = 0 |
| Case 5 | U1 = Free, U2 = Free, UR3 = 0 | U1 = 0, U2 = Free, UR3 = 0 | U1 = 0, U2 = 0, UR3 = 0 |
| Case 6 | U1 = 0, U2 = Free, UR3 = 0 | U1 = Free, U2= Free, UR3 = 0 | U1 = 0, U2 = 0, UR3 = 0 |
| Case 7 | U1 = 0, U2 = Free, UR3 = Free | U1 = 0, U2=Free, UR3 =0 | U1 = 0, U2= 0, UR3 = Free |
| Case 8 | U1 = 0, U2 = Free, UR3 = 0 | U1 = 0, U2 = Free, UR3 = Free | U1 = 0, U2= 0, UR3 = Free |
| Case 9 | U1 = 0, U2 = Free, UR3 = Free | U1 = 0, U2 = Free, UR3 = 0 | U1 = Free, U2 = 0, UR3 = 0 |
| Case 10 | U1 = 0, U2 = Free, UR3 = 0 | U1 = 0, U2 = Free, UR3 = Free | U1 = Free, U2 = 0, UR3 = 0 |
| Case 11 | U1 = Free, U2 = Free, UR3 = 0 | U1 = 0, U2 = Free, UR3 = 0 | U1 = Free, U2 = 0, UR3 = 0 |
| Case 12 | U1 = 0, U2 = Free, UR3 = 0 | U1 = Free, U2 = Free, UR3 = 0 | U1 =Free, U2=0, UR3 =0 |

**Table 2. Multiple Boundary conditions.** We use the model training on the strain trajectory of material geometry with specific boundary condition, to predict for unseen boundary condition cases. Different cases tested are provided in this table. U1 = x direction translation, U2 = y-direction translation and UR3 = z-direction rotation.

### 3.4 Super Resolution
#### 3.4.1 Pixel Wise Super-Resolution.

In the traditional FE analysis, the accuracy of results depends heavily on the size of the discretization used. A finer mesh is usually required to capture regions of high-stress concentration accurately, but this can significantly increase the computational cost and analysis time. NOs with their zero-shot super-resolution capabilities can address this issue. In this study, we evaluate NOs' super-resolution performance by using models trained on low-resolution data to make predictions for higher-resolution data. Specifically, we train our models on 48 × 48 domains and use them to output results for resolutions of 28 × 28 (Figure 9(a)), 104 × 104 (Figure 9(b)) and 200 × 200 (Figure 9c)) for t = 4 to 11 timesteps, which were not part of the training data. FNO is the only NO that is able to perform pixel-wise super-resolution, while MWT outputs unreliable results. This capability, FNO is highly desirable as it can be used to obtain high-fidelity, high-resolution data for multiple time steps outputted from a model trained on cheaper data. Such performance significantly reduces the costs of conducting high-precision analysis using ML-based surrogates.

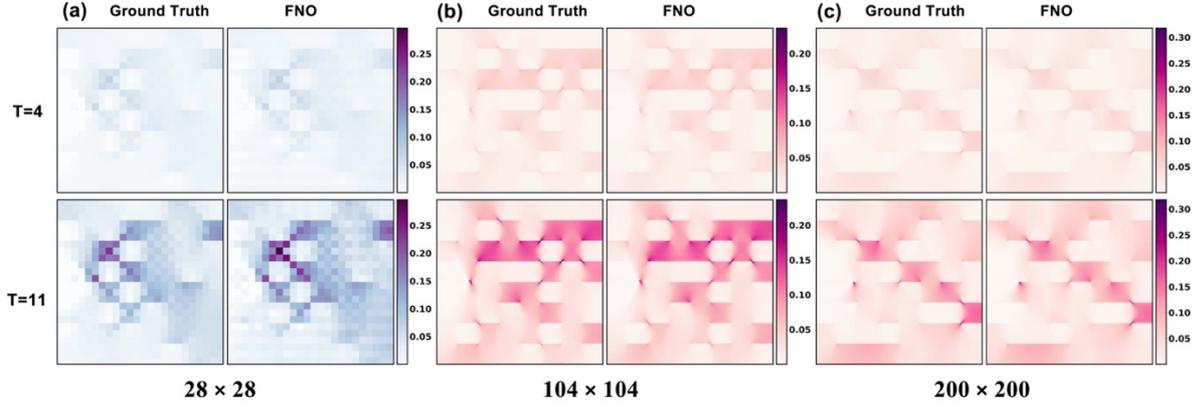

**Figure 9.** Pixel wise zero-shot super resolution. FNO trained on the spatial dimensions of 48×48 is evaluated on structures with dimensions of **(a)** 28 × 28 **(b)** 104 × 104 **(c)** 200 × 200.

### 3.4.2 Material Super Resolution

In order to obtain the strain trajectory, FE simulations are conducted on a 2D composite consisting of an 8×8 grid of soft and stiff units, with an overall image resolution of 48×48 for both input and output. While this resolution suffices for a simpler material configuration, real-world applications often involve more complex microstructures with higher material grid resolution. To address this, we leverage the super-resolution capabilities of NOs, which enable higher output resolution from models trained on lower-resolution data. Unlike classical solvers, which are sensitive to the size of discretization, NOs claim to transfer solutions from a lower to a higher resolution. By the design of architectures, only FNO and MWT are discretization invariant and can be tested on inputs sized differently than what they are actually trained on. In the case of material super-resolution, we test the models with a material grid size of 16×16 and an overall image resolution of 96×96. We present the results for FNO and MWT in Figure 10(a) depicting the strain maps for t=4,11 time step. The results show that FNO performs the best with the slightest error, whereas MWT fails to predict the correct output. Even though MWT shows better results for super-resolution while predicting the flow problems (G. Gupta et al., 2022), it suffers in capturing the strain outputs for a finer material grid input. While analyzing results for FNO, we observe that adding more Fourier layers to the FNO network works against the super-resolution capabilities. For example, on adding one extra Fourier and convolution layer, the model fails to predict for fine-resolution geometries. One of the reasons might be the over-smoothening due to the Fourier layers.

### 3.4.3 Arbitrary Geometry Cases

The microstructure geometry utilized in this study exhibits a checkerboard pattern consisting of soft and stiff units in equal proportions. However, real-world microstructures can possess far greater complexity and variability in terms of their designs and proportions of composite phases. Therefore, a deep learning model capable of generalizing to arbitrary geometries is highly desirable. We utilize FNO to predict the strain trajectory of an arbitrary material geometry not present in the training stage. Figure 10(b) displays the strain values at t = 4,11 for a typical microstructure. Based on our experiments with arbitrary geometries, we find that MWT is unable to accurately predict super-resolution for composites undergoing tension. This observation is consistent with the aforementioned results for material and pixel-wise super-resolution and indicates that MWT may not be suitable for predictions in such scenarios. It is noteworthy to mention that the accuracy achieved is lower than that of the predictions for checkerboard geometries. However, there have been no studies that have demonstrated super-resolution while predicting stress/strain values. The size of the solution image is 96×96, allowing for higher resolution predictions for previously unseen geometries.

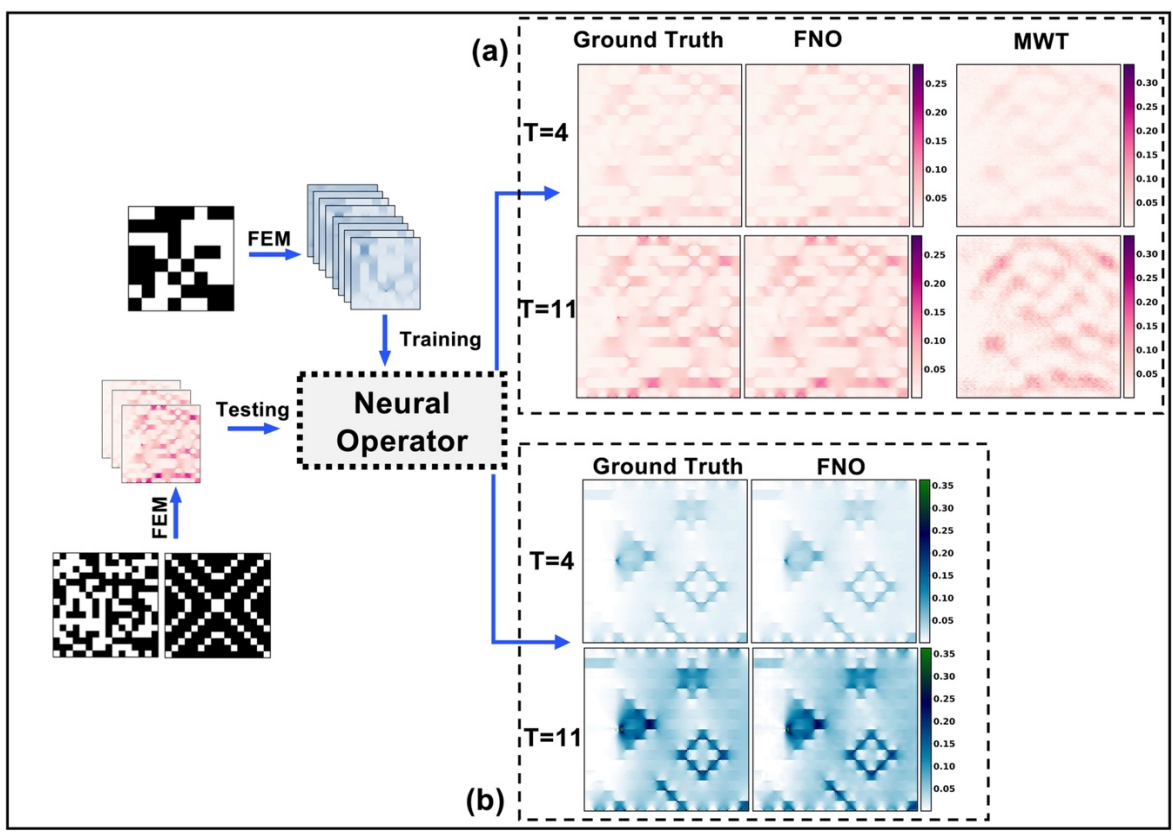

**Figure 10.** Zero shot generalization to unseen geometries. **(a)** The model trained on 8 × 8 material resolution is used to predict strains for geometries having 16× 16 material resolution. **(b)** The input geometries used to generate the ground truth had a checkerboard structure. Model trained on such data is used to estimate the strain values for unseen arbitrary geometries as shown in the figure.

## 4. Conclusion

In the present work, we critically evaluate the ability of NOs toward predicting the strain evolution in digital composites. We demonstrate that NOs can reliably predict the strain evolution in digital composites of arbitrary geometry from an extremely small number of training dataset (~1000 in this case). Further, the trained model exhibit reasonable prediction of the strain evolution until fracture. The models also exhibit the ability to predict the strain evolution for unseen boundary conditions making it an extremely useful and economic surrogate model instead of the FE simulations. Interestingly, FNO is the only NO that exhibits pixel and material super-resolutions. This unique property of FNO enables the surrogate model to provide high fidelity results at different length scales and resolutions different from that of the trained one. Altogether, the present work demonstrates that NOs can be efficient, economic, and reliable surrogate models that can either be used as a stand-alone predictor or in tandem with an FE model (where FE is used occasionally to correct the propagating errors of NO) to accelerate the modelling of complex digital composites. Such approaches can potentially enable the efficient design of composite materials for targeted applications by tailoring the microstructure.

## 5. Acknowledgments

The authors thank the High-Performance Computing (HPC) facility at IIT Delhi for computational and storage resources.

## 6. Code Availability

The codes used for this study are available at [M3RG-IITD/StrainEvolution (github.com)](github.com)

## 7. Author Contribution

SC and NMAK supervised the work. MMR performed simulations, trained models, post-processing, and initial draft. All the authors analysed the results and edited the manuscript.

**Reference:**

Barbero, E.J., 2017. Introduction to Composite Materials Design. CRC Press.

Bhaduri, A., Gupta, A., Graham-Brady, L., 2022. Stress field prediction in fiber-reinforced composite materials using a deep learning approach. Composites Part B: Engineering 238, 109879. https://doi.org/10.1016/j.compositesb.2022.109879

Butler, K.T., Davies, D.W., Cartwright, H., Isayev, O., Walsh, A., 2018. Machine learning for molecular and materials science. Nature 559, 547–555. https://doi.org/10.1038/s41586-018-0337-2

Cao, S., Wang, H., Lai, Y., Bo, R., Feng, X., 2023. Modeling the nonlinear responses of soft network materials assisted by masked-fusion artificial neural network. Materials Today Communications 35, 106013. https://doi.org/10.1016/j.mtcomm.2023.106013

Chawla, R., Sharma, S., 2018. A molecular dynamics study on efficient nanocomposite formation of styrene–butadiene rubber by incorporation of graphene. Graphene Technol 3, 25–33. https://doi.org/10.1007/s41127-018-0018-9

Chen, P.-Y., McKittrick, J., Meyers, M.A., 2012. Biological materials: Functional adaptations and bioinspired designs. Progress in Materials Science 57, 1492–1704. https://doi.org/10.1016/j.pmatsci.2012.03.001

Friemann, J., Dashtbozorg, B., Fagerström, M., Mirkhalaf, S.M., 2023. A micromechanics-based recurrent neural networks model for path-dependent cyclic deformation of short fiber composites. International Journal for Numerical Methods in Engineering 124, 2292–2314. https://doi.org/10.1002/nme.7211

Goswami, S., Yin, M., Yu, Y., Karniadakis, G.E., 2022. A physics-informed variational DeepONet for predicting crack path in quasi-brittle materials. Computer Methods in Applied Mechanics and Engineering 391, 114587. https://doi.org/10.1016/j.cma.2022.114587

Gu, G.X., Dimas, L., Qin, Z., Buehler, M.J., 2016. Optimization of Composite Fracture Properties: Method, Validation, and Applications. Journal of Applied Mechanics 83. https://doi.org/10.1115/1.4033381

Guo, K., Buehler, M.J., 2020. A semi-supervised approach to architected materials design using graph neural networks. Extreme Mechanics Letters 41, 101029. https://doi.org/10.1016/j.eml.2020.101029

Gupta, A., Bhaduri, A., Graham-Brady, L., 2022. Accelerated multiscale mechanics modeling in a deep learning framework. https://doi.org/10.48550/arXiv.2212.14601

Gupta, G., Xiao, X., Bogdan, P., 2022. Multiwavelet-based Operator Learning for Differential Equations. Presented at the Advances in Neural Information Processing Systems.

Haghighat, E., Raissi, M., Moure, A., Gomez, H., Juanes, R., 2021. A physics-informed deep learning framework for inversion and surrogate modeling in solid mechanics. Computer Methods in Applied Mechanics and Engineering 379, 113741. https://doi.org/10.1016/j.cma.2021.113741

Hasan, T., Capolungo, L., Zikry, M.A., 2023. Predictive machine learning approaches for the microstructural behavior of multiphase zirconium alloys. Sci Rep 13, 5394. https://doi.org/10.1038/s41598-023-32582-9

Henkes, A., Wessels, H., Mahnken, R., 2022. Physics informed neural networks for continuum micromechanics. Computer Methods in Applied Mechanics and Engineering 393, 114790. https://doi.org/10.1016/j.cma.2022.114790

Hughes, T.W., Williamson, I.A.D., Minkov, M., Fan, S., 2019. Wave physics as an analog recurrent neural network. Science Advances 5, eaay6946. https://doi.org/10.1126/sciadv.aay6946

Jensen, Z., Kim, E., Kwon, S., Gani, T.Z.H., Román-Leshkov, Y., Moliner, M., Corma, A., Olivetti, E., 2019. A Machine Learning Approach to Zeolite Synthesis Enabled by Automatic Literature Data Extraction. ACS Cent. Sci. 5, 892–899. https://doi.org/10.1021/acscentsci.9b00193

Kadic, M., Milton, G.W., van Hecke, M., Wegener, M., 2019. 3D metamaterials. Nat Rev Phys 1, 198–210. https://doi.org/10.1038/s42254-018-0018-y

Kairn, T., Daivis, P.J., Ivanov, I., Bhattacharya, S.N., 2005. Molecular-dynamics simulation of model polymer nanocomposite rheology and comparison with experiment. J. Chem. Phys. 123, 194905. https://doi.org/10.1063/1.2110047

Lamb, A., Goyal, A., Zhang, Y., Zhang, S., Courville, A., Bengio, Y., 2016. Professor Forcing: A New Algorithm for Training Recurrent Networks. https://doi.org/10.48550/arXiv.1610.09038

LeCun, Y., Bengio, Y., Hinton, G., 2015. Deep learning. Nature 521, 436–444. https://doi.org/10.1038/nature14539

Li, Z., Kovachki, N., Azizzadenesheli, K., Liu, B., Bhattacharya, K., Stuart, A., Anandkumar, A., 2021. Fourier Neural Operator for Parametric Partial Differential Equations (No. arXiv:2010.08895). arXiv. https://doi.org/10.48550/arXiv.2010.08895

Liu, Y., Zhang, X., 2011. Metamaterials: a new frontier of science and technology. Chem. Soc. Rev. 40, 2494–2507. https://doi.org/10.1039/C0CS00184H

Manual, A.U., 2014. Abaqus theory guide. Version 6, 281.

Maurizi, M., Gao, C., Berto, F., 2022. Predicting stress, strain and deformation fields in materials and structures with graph neural networks. Sci Rep 12, 21834. https://doi.org/10.1038/s41598-022-26424-3

Meza, L.R., Das, S., Greer, J.R., 2014. Strong, lightweight, and recoverable three-dimensional ceramic nanolattices. Science 345, 1322–1326. https://doi.org/10.1126/science.1255908

Mozaffar, M., Bostanabad, R., Chen, W., Ehmann, K., Cao, J., Bessa, M.A., 2019. Deep learning predicts path-dependent plasticity. Proceedings of the National Academy of Sciences 116, 26414–26420. https://doi.org/10.1073/pnas.1911815116

Nie, Z., Jiang, H., Kara, L.B., 2019. Stress Field Prediction in Cantilevered Structures Using Convolutional Neural Networks. Journal of Computing and Information Science in Engineering 20. https://doi.org/10.1115/1.4044097

Oommen, V., Shukla, K., Goswami, S., Dingreville, R., Karniadakis, G.E., 2022. Learning two-phase microstructure evolution using neural operators and autoencoder architectures. npj Comput Mater 8, 1–13. https://doi.org/10.1038/s41524-022-00876-7

Paszke, A., Gross, S., Massa, F., Lerer, A., Bradbury, J., Chanan, G., Killeen, T., Lin, Z., Gimelshein, N., Antiga, L., Desmaison, A., Köpf, A., Yang, E., DeVito, Z., Raison, M., Tejani, A., Chilamkurthy, S., Steiner, B., Fang, L., Bai, J., Chintala, S., 2019. PyTorch: An Imperative Style, High-Performance Deep Learning Library. https://doi.org/10.48550/arXiv.1912.01703

Pham, M.-S., Liu, C., Todd, I., Lertthanasarn, J., 2019. Damage-tolerant architected materials inspired by crystal microstructure. Nature 565, 305–311. https://doi.org/10.1038/s41586-018-0850-3

Qin, Z., Yu, Q., Buehler, M.J., 2020. Machine learning model for fast prediction of the natural frequencies of protein molecules. RSC Adv. 10, 16607–16615. https://doi.org/10.1039/C9RA04186A

Raissi, M., Perdikaris, P., Karniadakis, G.E., 2019. Physics-informed neural networks: A deep learning framework for solving forward and inverse problems involving nonlinear partial differential equations. Journal of Computational Physics 378, 686–707. https://doi.org/10.1016/j.jcp.2018.10.045

Rashid, M.M., Pittie, T., Chakraborty, S., Krishnan, N.M.A., 2022. Learning the stress-strain fields in digital composites using Fourier neural operator. iScience 25, 105452. https://doi.org/10.1016/j.isci.2022.105452

Su, I., Jung, G.S., Narayanan, N., Buehler, M.J., 2020. Perspectives on three-dimensional printing of self-assembling materials and structures. Current Opinion in Biomedical Engineering, Biomechanics and Mechanobiology: Growth and remodeling in both mechanics and mechanobiology 15, 59–67. https://doi.org/10.1016/j.cobme.2020.01.003

Tripura, T., Chakraborty, S., 2023. Wavelet Neural Operator for solving parametric partial differential equations in computational mechanics problems. Computer Methods in Applied Mechanics and Engineering 404, 115783. https://doi.org/10.1016/j.cma.2022.115783

Wegst, U.G.K., Bai, H., Saiz, E., Tomsia, A.P., Ritchie, R.O., 2015. Bioinspired structural materials. Nature Mater 14, 23–36. https://doi.org/10.1038/nmat4089

Xue, T., Adriaenssens, S., Mao, S., 2023. Learning the nonlinear dynamics of mechanical metamaterials with graph networks. International Journal of Mechanical Sciences 238, 107835. https://doi.org/10.1016/j.ijmecsci.2022.107835

Yang, C., Kim, Y., Ryu, S., Gu, G.X., 2020. Prediction of composite microstructure stress-strain curves using convolutional neural networks. Materials & Design 189, 108509. https://doi.org/10.1016/j.matdes.2020.108509

Yang, Z., Yu, C.-H., Buehler, M.J., 2021a. Deep learning model to predict complex stress and strain fields in hierarchical composites. Science Advances 7, eabd7416. https://doi.org/10.1126/sciadv.abd7416

Yang, Z., Yu, C.-H., Guo, K., Buehler, M.J., 2021b. End-to-end deep learning method to predict complete strain and stress tensors for complex hierarchical composite microstructures. Journal of the Mechanics and Physics of Solids 154, 104506. https://doi.org/10.1016/j.jmps.2021.104506

Zhang, E., Dao, M., Karniadakis, G.E., Suresh, S., 2022. Analyses of internal structures and defects in materials using physics-informed neural networks. Science Advances 8, eabk0644. https://doi.org/10.1126/sciadv.abk0644

Zhang, X., Wang, Y., Ding, B., Li, X., 2020. Design, Fabrication, and Mechanics of 3D Micro-/Nanolattices. Small 16, 1902842. https://doi.org/10.1002/smll.201902842